\documentclass{elsart}
\usepackage{natbib}

\usepackage{epsf,graphicx}

\newcommand{\eg}{{\em e.g., }}           
\newcommand{\ie}{{\em i.e., }}           

\def\be{\begin{equation}}
\def\ee{\end{equation}}
\def\ba{\begin{array}}
\def\ea{\end{array}}
\def\Keff{K_{\rm eff}}
\usepackage{amssymb,amsmath}  
\newcommand{\T}[1]{{\boldsymbol{#1}}}

\begin{document}
\begin{frontmatter}

\title{Effective toughness of\\ heterogeneous brittle materials}

\author[SVI]{S. Roux,}
\author[SVI]{D. Vandembroucq}
\author[LMT]{and F. Hild}

\address[SVI]{Laboratoire Surface du Verre et Interfaces,\\
Unit\'{e} Mixte de Recherche CNRS/Saint-Gobain,\\
39 Quai Lucien Lefranc, 93303 Aubervilliers Cedex, France.}
\address[LMT]{LMT-Cachan,\\
ENS de Cachan / CNRS-UMR 8535 / University Paris 6,\\
61 Avenue du Pr\'esident Wilson, 94235 Cachan Cedex, France.}

\begin{abstract}
A heterogeneous brittle material characterized by a random field
of local toughness $K_c(\T{x})$  can be represented by an
equivalent homogeneous medium of toughness, $\Keff$.
Homogenization refers to a process of estimating $\Keff$ from the
local field $K_c(\T{x})$.  An approach based on a perturbative
expansion of the stress intensity factor along a rough crack front
shows the occurrence of different regimes depending on the
correlation length of the local toughness field in the direction
of crack propagation.  A {\sl ``weak pinning''} regime takes place
for long correlation lengths, where the effective toughness is the
average of the local toughness.  For shorter correlation lengths,
a transition to {\sl ``strong pinning''} occurs leading to a much
higher effective toughness, and characterized by a propagation
regime consisting in jumps between pinning configurations.
\end{abstract}


\end{frontmatter}

\maketitle

\section{Introduction}

Brittle materials such as ceramics, glass and rocks are known to
be extremely sensitive to bulk and surface defects, from which
cracks can be initiated eventually leading to failure. This
extreme sensitivity calls for a statistical analysis of crack
initiation, which has been extensively developed following the
pioneering work of Weibull \citep{Weibull39,Freudenthal68}. This
approach, describing the inception of crack propagation caused by
initial defects \citep{jayatilaka77}, has been progressively
extended to account for various statistical distributions of bulk
or surface defects \citep{munz99}, multiaxial criteria for
critical loads on defects \citep{batdorf78,evans78}, inhomogeneous
stress fields \citep{davies73}. In this ``weakest-link'' approach,
the analysis is focused on initiation, implicitly assuming that
the propagation stage is obtained systematically over unlimited
distances, due to the lower loading needed for this stage as
compared to the initiation one.

However, in some cases, even though cracks have been nucleated,
they will not propagate to large distances, and their presence may
still be acceptable in service condition of a given structure.
Simple examples of such confined cracks are those induced by
indentation \citep{lawn93}. In the latter case, the stress field
has a rapid decay with distance from the indentation point, and
hence a crack which can easily be nucleated may stop shortly after
initiation. In ceramic~/~metal assemblies, residual stresses
caused by the coefficient of thermal expansion mismatch can also
prevent cracks to traverse the brittle part so that a weakest link
hypothesis does not apply \citep{charles02}. In brittle-matrix
composites, crack arrest is also observed due to the bridging
forces induced by the fibers \citep{evans90}.

Such situations require the characterization of the conditions
under which a ``macroscopic'' crack may or may not propagate. The
word ``macroscopic'' requires a specific attention.  At a
microscopic scale, $\lambda$, a number of non-linear phenomena may
take place in the so-called process-zone (\eg dislocation
emission, damage initiation, or simply non-linear debonding).
However, at a larger scale, the effect of all those confined
non-linearities can be characterized by a toughness $K_c$ which
will dictate whether a crack of stress intensity factor $K$ can
(when $K \geq K_c$) or cannot (when $K<K_c$) propagate. The
situation considered in the present study is when, at a scale
larger than $\lambda$, the toughness may vary from places to
places, $K_c(\T{x})$ thus being a random field whose
characteristics are:

\begin{itemize}
 \item
a probability distribution function $p(K_c)$;\\

 \item
a correlation function $C$ defined as
\begin{equation}
C(\T{x})=\langle K_c(\T{y})K_c(\T{x}+\T{y})\rangle_{\T{y}} -\langle K_c(\T{y})
\rangle_{\T{y}}^2 \,,
\end{equation}

where the brackets $\langle ... \rangle_{\T{y}}$ denote an average
over the coordinate $\T{y}$.

\end{itemize}

Let us focus in the following on cases where the correlation
length $\xi$ (above which $C(x) \approx 0$) can be defined, at
least along the crack front.  The term ``macroscopic'' refers
specifically to the case of a crack whose front length $L$ is much
greater than $\xi_x$ along the direction of the crack front.  In
such a situation, the medium can be characterized by an effective
toughness that controls the propagation or arrest of the crack.
More precisely, the local crack front roughness caused by the
random toughness landscape can be ignored, and an equivalent crack
having a straight front and the same mean position is defined.
Along this equivalent crack, a loading leading to a constant
stress intensity factor (SIF) can be considered and the same
loading will be applied simultaneously on the heterogeneous
material. The critical SIF computed on the equivalent geometry
which corresponds to the onset of propagation will thus
characterize the effective toughness, $\Keff$.

The problem of a crack propagating in a heterogeneous toughness
field has been considered both numerically and experimentally, for
two-phase materials having well defined geometries.  In the
context of a layered toughness along the mean crack front,
Eriksson \citep{Eriksson98} used an energy based argument to
estimate the effective toughness by a rule of mixture. The
by-passing of a tough inclusion has been simulated numerically in
great details \citep{bower91,bower93} and corresponding
experiments have been performed \citep{experiment}. Curtin
\citep{curtin97,curtin98} performed a simplified analysis of a
similar problem. To analyze the crack advance in fiber reinforced
composites, instead of toughness distributions, strength
distributions have been used
\citep{beyerlein97a,beyerlein97b,curtin98,landis00}. However the
question of the large scale effective toughness was not
considered.  The relaxation of a crack distorsion due to an
obstacle in a dynamic situation has also been considered in
\citep{Rice00,Rice01,Woolfries99}, with a statistical treatment
addressing the question of crack front roughening.

The heterogeneity of the toughness field induces perturbations of
the crack front geometry. A large amount of work has been carried
out in recent years to estimate the effect of a rough crack
geometry on the stress intensity factor at the crack front in
different cases (planar or three-dimensional, static or dynamic,
...) \citep{GaoRice89,Movchan95,Willis95,Movchan98}.  In the
restricted context of a crack propagating in anti-plane geometry
(mode III), Vandembroucq and Roux~\citep{VR97JMPS} performed a
second order expansion of the stress intensity factor in crack
roughness. The second order term was shown to systematically
decrease the stress intensity factor compared with the expected
value for a straight crack. This was interpreted as a
``strengthening'' effect, \ie an increase of the apparent
toughness of the material.

Alternatively, a statistical physics approach of crack front
pinning by a random field is proposed by different authors mostly
from a theoretical perspective
\citep{Perrin94,schmittbuhl95,Bouchaud1,Ramanathan97,Ramanathan98,Tanguy98,Bouchaud2,Hansen03}
accompanied by some experimental investigations
\citep{Schmittbuhl97}. Most of these studies, however, focus on
the statistical features of crack front roughness, and interesting
features attached to the onset of crack propagation interpreted as
a depinning transition. However, little attention has been paid to
the quantitative estimate of the effective toughness. In
\citep{Skoe02}, the question of the statistical distribution of
the macroscopic SIF at the onset of propagation is addressed,
again revealing a universal critical behavior close to an
effective threshold. However, the quantitative value of the latter
was not treated.

Crack arrest conditions have also been studied when the crack tip
traverses a medium with random toughness. The problem to solve
concerns a solid medium consisting of elastic brittle grains or
potential arrest sites. The grain size is considered to be of
constant size and toughness is constant in each grain. The
toughness varies from grain to grain so that no spatial
correlations on scales larger than the grain size are assumed
\citep{chudnovsky87}. The same hypothesis was used by Charles et
al. \citep{charles02,charles03} to analyze instantaneous and
delayed crack propagation and arrest. Jeulin \citep{jeulin94}
proposed a model in which the microstructure is assumed to be
described by a Poisson mosaic. A Poisson tessellation defines the
grain boundaries. The latter are made of Poisson lines in the
plane for a two-dimensional medium. Instantaneous propagation and
arrest conditions are investigated.

In the following, the case of a layered local toughness, invariant
along the direction of propagation, is considered first, since
this particular texture allows for a closed-form answer.  Then
slow modulations of the toughness along the propagation direction
are considered, in the so-called ``weak pinning'' regime. Finally,
the ``strong pinning'' case is addressed when the correlation
length of the toughness perpendicular to the front becomes small.
In all these cases, an emphasis is put on an estimate of the
effective toughness and comments are made on specific features
expected in the propagation regime.

\section{Layered toughness}

A crack front parallel to the $x$-axis is considered, with a
toughness field translationally invariant along the propagation
direction $y$, $\partial K_c(x,y)/\partial y=0$.  Starting from a
straight front, as the loading is increased, the crack front
begins to develop some roughness, being still pinned in some
regions and propagating in other parts.  The front is
characterized by its coordinate along the $y$ axis: $h(x)$.  For
simplicity, we assume that the crack remains planar, confined into
the $(x,y)$-plane.

\begin{figure}
\centerline{\epsfxsize=.35\hsize \epsffile{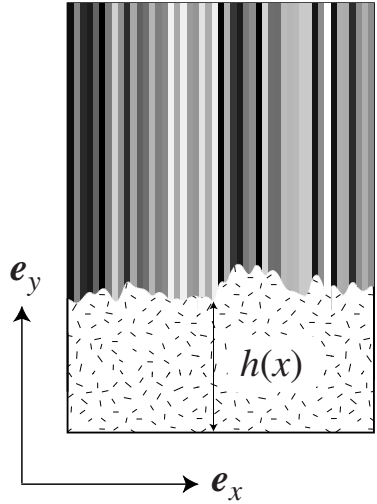}}
\caption{\label{fi:layered} Crack propagation in a layered
toughness configuration. The crack front is distorted such that at
every point $x$ along the front the stress intensity factor $K(x)$
matches exactly the local toughness value $K_c(x)$.}
\end{figure}

The roughness of the crack front will induce a modulation of the
local SIF, $K(x)$, when compared to the macroscopic one $K_0$,
defined with the same loading and a straight crack having the same
mean position.  By using a perturbation analysis
\citep{GaoRice89}, a first order solution can be derived
 \begin{equation}
\label{eq1}
 K(x)=K_0\left(1+{1\over\pi}\int \frac{h(x')-h(x)}{(x'-x)^2}
 dx'\right) \,.
 \end{equation}
At the onset of propagation, the stress intensity factor has to
match the local toughness (\ie $K(x)=K_c(x)$).  This can be
achieved through a particular conformation of the crack front,
$h^*(x)$, which can be obtained from a mere Fourier transform of
Eq.~(\ref{eq1}). By denoting Fourier transforms with a
$\widetilde{~}$ sign,
\begin{equation}
\label{eq:Fourier}
  \widetilde{h^*}(k)=\frac{1}{|k|}\frac{\widetilde{K_c}(k)}{K_0} \,.
\end{equation}
The integration of Eq.~(\ref{eq1}) over $x$, using $K(x)=K_c(x)$,
allows one to derive the effective toughness with $\Keff=K_0$
 \begin{equation}
\label{eq:exacte}
 \Keff=\langle K_c\rangle \,.
 \end{equation}
Equation~(\ref{eq:exacte}) is an exact result, however limited to
the case of a narrow distribution of local toughness so that the
first order expansion~(\ref{eq1}) of the SIF modulation remains
valid.  We note that this simple conclusion is a general result
when using a first order coupling term between different positions
along the front.  We note that, still using a similar linear
kernel, a following section will show that a regime (termed
``strong-pinning'') may appear where such a simple conclusion
breaks down.


\section{Self-consistent homogenization}

Let us now consider the more general case of a random toughness
field, varying along the propagation direction.  In contrast to
the layered case, an exact result is not derived, but rather an
approach based on a self-consistent approximation is proposed,
similar to the one used for homogenization of a randomly
heterogeneous elastic solid~\citep{kroener67}.

\begin{figure}
\centerline{\epsfxsize=.75\hsize \epsffile{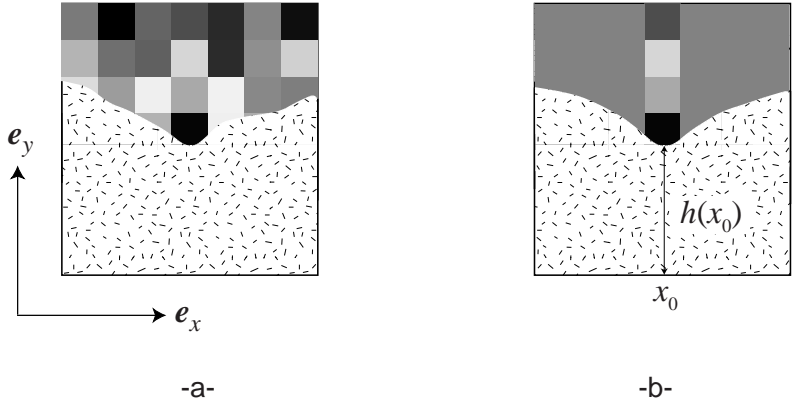}}
\caption{\label{fi:random_w} A random toughness configuration is
represented schematically on the left (a) with a grey-scale coded
toughness.  On the right, (b), a reference case is constructed by
extracting the toughness pattern along a line parallel to the
propagation direction, and substituting to the rest a uniform
toughness environment. The latter is determined such that the
deflection $\Delta h = h(x_o)- \langle h(x) \rangle_x$ is zero on
average over all accessible crack front conformations.}
\end{figure}

A narrow strip is considered parallel to the propagation direction
$y$, centered on $x_0$, and of width $\xi_x$ equal to the
correlation length of the toughness field in the $x$-direction.
Within this strip, the random toughness is preserved. Outside it,
one substitutes to the random toughness a homogeneous toughness,
$K_0$ as shown schematically in Figure~\ref{fi:random_w}.
The local toughness contrast induces a perturbation of the
equilibrium position of the crack front. Depending on the sign of
the toughness contrast, the position $h(x_0)$ of the front at the
center of the strip is either in front of or behind the average
front position $\langle h(x) \rangle_x$. The value of this
deflection $\Delta h(x_0) = h(x_0)- \langle h(x) \rangle_x$ is at
first order directly proportional to the toughness difference
$\Delta K(x_0)= K(x_0)-K_0$. In the process of homogenization, the
value of $K_0$ is chosen such that $\langle \Delta h(x_0)
\rangle_x$ vanishes, \ie the front shape averaged over all crack
front conformations along the disordered strip is flat.

Such an approach turns out to provide rather accurate estimates in
the case of disordered elastic solids \citep{Willis91}, albeit
there is no way to estimate {\it a priori} the accuracy of the
result.

Equation~(\ref{eq1}) can be used to compute the crack front shape.
The stress intensity factor to consider is $K=K_0$ all along the
front except in the strip where $K=K(x_0,h(x_0))$. Inversion of
Eq.~(\ref{eq:Fourier}) leads to a logarithmic front shape, and
thus the extension of the front along the $x$-direction, $L$,
always comes into play in the difference $\Delta h=h(x_0)-\langle
h\rangle$.  Homogeneity also dictates that the system size has to
be scaled by the strip width, $\xi_x$. Because of linearity,
$\Delta h$ is proportional to $K(x_0,h(x_0))-K_0$ so that
 \begin{equation}
\label{eq:spring}
 K(x_0,\langle h\rangle +\Delta h)-K_0=-
 \frac{A K_0}{\xi_x\log(L/\xi_x)}\Delta h \,,
 \end{equation}
where $A$ is a numerical constant.

\begin{figure}
\centerline{\epsfxsize=.8\hsize \epsffile{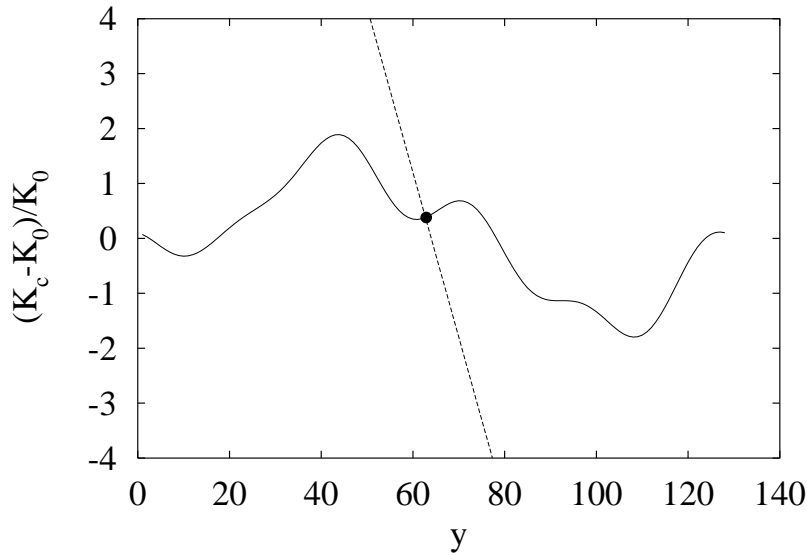}}
\caption{\label{fi:weakpin} Schematic construction of the
equilibrium position of the crack front shape: the intersection of
the toughness profile shown as a continuous curve, with the
coupling due to the distorsion of the crack front shown as a
dotted line gives here a unique equilibrium position shown as a
black dot.}
\end{figure}

Using the linear dependence between the stress intensity factor
and the distance between the crack front within the strip and away
from it, the equilibrium front position can be determined as shown
in Fig.~\ref{fi:weakpin}, by plotting $K_c(x_0,\langle h\rangle
+\Delta h)$ as a function of $\Delta h$ together with the linear
relation (see Eq.~(\ref{eq:spring})).

\section{Weak pinning regime}

When the correlation length along the $y$-direction is long, so
that the gradient $\partial K(x,y)/\partial y$ remains always
smaller than the ``stiffness'' $S=A K_0/[\xi_x\log(L/\xi_x)]$, a
regime referred to as ``weak pinning'' occurs. In this case, there
is a unique equilibrium position for the crack front
(Fig.~\ref{fi:weakpin}) since the ``stiffness'', $S$, is greater
than the gradient of toughness in the propagation ($y$-)direction.
To compute the effective toughness $\Keff=K_0$, one has to
integrate the $K_c$ values with a measure corresponding to a
uniform sampling of $\langle h\rangle$.  This introduces a bias in
the weighting of the distribution of $K_c$ values
 \begin{eqnarray}
 \nonumber
 \Keff&=&\lim_{Y\to\infty}\frac{1}{Y}\int_0^Y K_c(y) \frac{d(\Delta h)}{dy}  dy \\
 &=&\lim_{Y\to\infty}\frac{1}{Y}\int_0^Y K_c(y) \left(1+
 \frac{1}{S}\frac{dK_c(y)}{dy}\right) dy \,.
 \end{eqnarray}
The first term gives the simple arithmetic average of the local
toughness, similar to the result obtained in the layered case. The
second term is proportional to the correlation between the local
toughness and its gradient.  By performing the change of
integration variable from $y$ to $K_c^2$, it becomes apparent that
in the case of a stationary toughness field, this second integral
vanishes.  Thus, in the weak-pinning regime, the same result as
for the layered case applies
 \begin{equation}
 \Keff=\langle K_c\rangle \,.
 \end{equation}
Therefore this simple result appears to be quite robust, and gives
confidence to the self-consistent approach proposed herein.

In this scenario, the propagation is smooth, \ie the local crack
front advance is a continuous function of the mean crack position.
The final result concerning the effective toughness could also
have been derived using different approaches, such as a global
energy balance \citep{Eriksson98}, resorting to the energy release
rate rather that the crack toughness.  We however refrain from
using such a treatment because of the following section, where
this argument will be shown to break down.

\section{Strong pinning regime}

\begin{figure}
\centerline{\epsfxsize=.8\hsize \epsffile{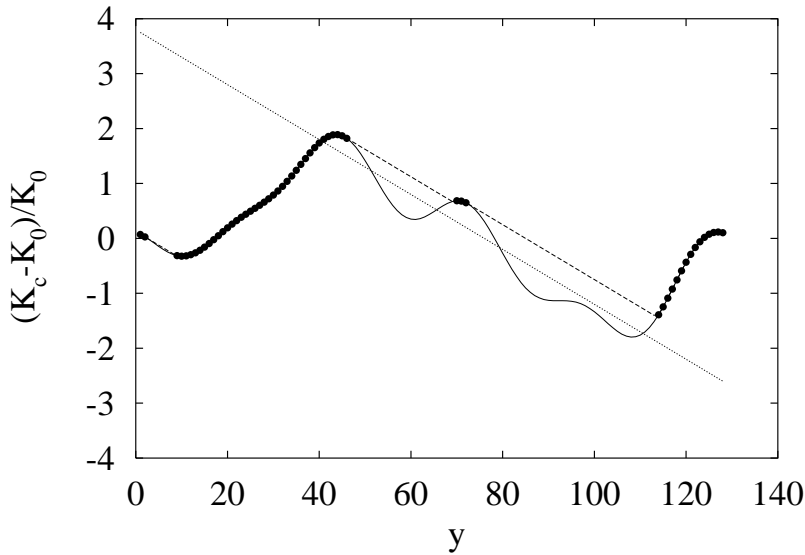}}
\caption{\label{fi:strongpin}
 Schematic construction for the strong pinning case.  The
continuous curve shows the local toughness profile. In contrast to
the previous figure, many solutions of equilibrium position exists
as shown by the five intersections of the dotted line with the
continuous curve (three stable positions and two unstable ones).
This selects only a subset of accessible arrest sites shown by
bold dots, separated by sudden jumps shown by a dashed line of
slope $-S$.}
\end{figure}

From the simple geometric construction leading to the equilibrium
position of the crack front, one sees that a very different
behavior appears as soon as the ``stiffness'', $S$, becomes
smaller than the gradient of toughness in the propagation ($y$-)
direction.  In this case, illustrated in Fig.~\ref{fi:strongpin},
multiple solutions can be found.  In generic cases, a sequence of
alternatively stable and unstable solutions is obtained.  However,
even ignoring those unstable positions,  the multiplicity of
solutions indicates that the position will be selected through the
history of propagation. In the case of interest, the mean crack
front position is chosen to be monotonically increasing, so that
the toughness values sampled in the strip are confined to the
highest values, as if the toughness profile were illuminated by a
grazing incidence light (\ie exposed hill tops correspond to
sampled equilibrium position, whereas shadowed regions correspond
to unstable jumps to a new arrest position).  In contrast to the
weak-pinning case where the propagation was smooth, one observes
in the strong-pinning case a sort of stick-slip crack propagation.
This regime is the one addressed in the statistical approaches
developed in
\citep{Bouchaud1,Ramanathan97,Ramanathan98,Tanguy98,Bouchaud2,Hansen03}.

To estimate $\Keff$, it is necessary to evaluate the probability
that a particular value of the toughness $K_c(x)$ can constitute
an accessible equilibrium position.  This implies that for all
$x'<x$, $K_c(x')<K_c(x)+S(x-x')$. To proceed, it is convenient to
specialize the toughness profile to a simple pattern, such as a
piecewise constant toughness over intervals of equal size $\xi_y$
(chosen equal to correlation length along the $y$ direction), and
without correlation from one interval to the next, as shown
schematically in Figure~(\ref{fi:random_w}a).

 Let $P(K_c)$ be
the cumulative toughness distribution $P(K)=\int_0^K p(k) dk$. The
probability, $f(K_c)$, that $K_c$ can be reached is written
 \begin{equation}
 \ba{rl}\label{eq:shadow}
 f(K_c)&\displaystyle=p(K_c)\prod_{i=1}^\infty P(K_c+iS\xi_y)\\
 &\displaystyle\approx p(K_c)\exp\left(1/(S\xi_y)\int_0^\infty
 \log(P(K_c+x)) dx\right)\,.
 \ea
 \end{equation}
This distribution is not normalized, as can be seen by the
existence of shadow zones containing unaccessible sites. Thus, the
p.d.f. of effective stable and accessible positions is
 \begin{equation}
 \phi(K_c)= \frac{f(K_c)}{\int_0^\infty f(k) dk}
 \end{equation}
and thus the effective toughness is again the simple arithmetic
average of $K_c$ using the probability distribution
 $\phi$ instead of $p$
 \begin{equation}
 \label{eq:final}
 \Keff= \frac{\int_0^\infty f(k) k~dk}{\int_0^\infty f(k)~dk}\,.
 \end{equation}
This last equation, together with the definition of $f$ in
Eq.~(\ref{eq:shadow}), gives an estimate of the effective
toughness based on the self-consistent approach.  Even though it
was derived in the context of strong-pinning, in the weak pinning
regime, Eq.~(\ref{eq:shadow}) provides $f(k)=p(k)$, and hence the
previously derived result can be recovered.

It is worth noting that the correlations along the crack front and
perpendicular to it contribute significantly to the final formula.
Consequently, the effective toughness is not a specific property
of the material along the potential crack plane, but may depend on
the direction of crack propagation.  This result contrasts with
energy-dissipation based arguments where such an orientation
effect cannot appear. The fundamental basis for such a difference
comes from the specificity of the strong pinning regime, namely
the occurrence of sudden jumps over unstable configurations.
During those jumps, the potential energy is transferred to kinetic
energy, and this dynamic aspect is not considered (\ie the
implicit assumption is that most of this kinetic energy is either
dissipated or radiated away from the crack so that it is no longer
available for propagation). This effect explains why energy-based
approaches will not reproduce the proposed result. Furthermore,
the crack length comes into play in the expression of $\Keff$ in
the strong-pinning regime through the particular dependence of the
stress intensity factor on the crack front distorsion.  However,
this dependence is only logarithmic and might be difficult to
observe.

Finally, the emphasis put on the multistability of
the strong pinning regime is comparable to parallel analyses
performed in other contexts such as solid friction
\citep{Caroli96}, wetting phenomena (hysteresis of wetting angle)
\citep{wetting}, or charge density waves \citep{Fisher} where
strong pinning is at the origin of new phenomena at the
macroscopic scale as compared to the microscopic one, and where
equilibrium concepts and energy balance arguments are no longer
operational.

\section{Summary}

In this article, the homogenization of a random toughness field is
addressed, taking into account the effect of the crack front
roughness on the local stress intensity factor to first order in
perturbation. Starting from a layered case, where it can be shown
that the effective toughness is equal to the arithmetic average of
the toughness, a self-consistent scheme is introduced to deal with
more general toughness fields.  This approach allows one to extend the
validity of the effective toughness estimate to slowly varying
toughness fields along the propagation direction, a situation
referred to as weak-pinning.  However, as the correlation length
of the toughness field along the propagation direction decreases,
a novel behavior is encountered when the crack front position is
no longer single-valued.  In the so-called strong-pinning regime, the
sampling of the local toughness becomes inhomogeneous, and gives
rise to an apparent strengthening.
The transition between these two regimes is progressive, and is
accompanied at the microscopic level by an unsteady propagation
displaying sudden jumps or burst similar to stick-slip.

\end{document}